\documentclass{aa}
\usepackage{graphics}
\begin{document}
\thesaurus{02(12.03.1)} 

\title{Analysis of CMB maps with 2D Wavelets}

\author{J. L. Sanz\inst{1} \and R. B. Barreiro\inst{1,2} \and 
L. Cay{\'o}n\inst{1} \and E. Mart{\'\i} nez-Gonz{\'a}lez\inst{1} \and
G. A. Ruiz\inst{3} \and F. J. D{\'\i} az\inst{3} \and
F. Arg{\"u}eso\inst{4} \and J. Silk\inst{5}  
\and L. Toffolatti\inst{6,7}}

\offprints{L.Cay{\'o}n}
\institute{Instituto de F{\'\i} sica de Cantabria, Fac. Ciencias, 
Avda. de los Castros s/n, 39005 Santander, Spain \and Dpto. F{\'\i} sica
Moderna, Universidad de Cantabria, Avda. de los Castros s/n, 
39005 Santander, Spain 
\and Dpto. de Electr{\'o}nica y Computadores, Universidad de Cantabria, 
Avda. de los Castros s/n, 39005 Santander,
Spain \and Dpto. Matem{\'a}ticas, Universidad de Oviedo, c/ Calvo Sotelo s/n, 
33007 Oviedo, Spain \and 
Astronomy Dept. \& Center for Particle Astrophysics, University of 
California, 601 Campbell Hall, Berkeley, CA 94720, USA \and Dpto. de 
F{\'\i} sica, Universidad de 
Oviedo, c/ Calvo Sotelo s/n, 33007 Oviedo, Spain \and Osservatorio Astronomico
di Padova, vicolo dell'osservatorio n5, 35122 Padova, Italy}

\date{Received ?? / Accepted ??}

\maketitle

\begin{abstract}
 
        We consider the 2D wavelet transform with two scales to study sky 
maps of temperature anisotropies in the cosmic microwave background 
radiation (CMB). 
We apply this technique to simulated maps of small sky patches 
of size 
%\simeq 
$12.8^{\circ}\times 12.8^{\circ}$ 
and $1.5^{\prime }\times 1.5^{\prime }$ pixels. 
The relation to the standard approach, based on the $C_l's$, is established 
through the introduction of the scalogram. We consider 
temperature fluctuations derived from standard, open and flat-$\Lambda$ 
Cold Dark Matter (CDM) models. 
We analyze CMB anisotropies maps plus uncorrelated Gaussian noise (uniform
and non-uniform) at different $S/N$ levels.
We explore in detail 
the denoising of such maps and compare the results with other techniques 
already proposed in the literature. Wavelet methods provide a good 
reconstruction of the image and power spectrum. Moreover, they are faster 
than previously proposed methods.

\end{abstract}

\section {Introduction}

Future Cosmic Microwave Background (CMB) experiments will provide high
resolution sky maps covering a wide range of frequencies. In addition to the 
cosmological CMB signal those maps will contain instrumental noise
and contributions from Galactic and extragalactic foregrounds. The
denoising of these maps as well as the separation of the different
components from the CMB signal are the most challenging problems for 
CMB cosmology. The final goal would be to reconstruct 
CMB maps trying not to loose structural details as well as to recover
the radiation power spectrum with the minimum error. 
In a first approach to these problems we present in this paper a 
denoising technique based on wavelets. Previously there have been other
works based in the use of Wiener filter (Tegmark and Efstathiou 
\cite{tegmark}) 
and Maximum Entropy Methods (Hobson et al. \cite{hobson98a,hobson99}). 
The use of 
denoising methods based on wavelets have certain advantages as
providing information of the contribution of different scales,
being computationally faster (0(N)) and not requiring iterative 
processes.
%and being not dependent on tuning parameters. 
The analysis of
discrete 2-dimensional images with wavelets can be performed 
following different approaches. The two computationally faster 
algorithms are the ones based on 
Multiresolution analysis (Mallat \cite{mallat}) and on 2D wavelet analysis
(Lemari{\'e} and Meyer \cite{lemarie}), using 
tensor products of one dimensional wavelets. A study of denoising of
CMB maps using the former method has been presented in Sanz et al. 
(\cite{sanz}).
This method is based on a single scale and three `details' at each
resolution level.
The 2-D wavelet method used in this work is based on two scales,
providing therefore more information on different resolutions (defined 
by the product of the two scales) than
the Multiresolution one. Moreover this technique is adapted to separable
wavelets. On the other hand, an analysis of denoising using 
spherical wavelets has been recently carried out by
Tenorio et al. (\cite{tenorio}).

The paper is organized as follows. In $\S2$ we present
some basic ideas about the continuous 2D wavelet
transform. We apply 2-D wavelets to the analysis of discrete 2D
CMB anisotropy maps of small sky patches in $\S3$ 
and the conclusions are presented in $\S4$.

\section{2D Continuous wavelet transform}

This section is dedicated to present the continuous form of the
2D wavelet approach we are later using to analyse discrete CMB maps.
As a difference with the multiresolution approach used in 
Sanz et al. (\cite{sanz}) the 2D wavelet method provides information
on many more resolution elements than the former method. Moreover, this
property is crucial for preforming an efficient linear denoising 
preserving the Gaussianity of the underlying CMB field (as will be
discussed in $\S3.2$).

The continuous wavelet transform of a 2D signal 
$f(x_1, x_2)$ is defined as\\
\begin{eqnarray}
w(R_1, R_2, b_1, b_2) & = & \int dx_1\,dx_2\,f(x_1, x_2) \nonumber \\
& & \times {\Psi } (R_1, R_2, b_1, b_2; x_1, x_2) \, ,
\end{eqnarray}
\begin{eqnarray}
{\Psi }(R_1, R_2, b_1, b_2; x_1, x_2) & = & 
\frac{1}{\sqrt{|R_1R_2|}} \nonumber \\
& & \times {\psi }
\left(\frac{x_1 - b_1}{R_1}, \frac{x_2 - b_2}{R_2}\right) \, ,
\end{eqnarray}
 
\noindent where $w(R_1, R_2, b_1, b_2)$ is the wavelet coefficient associated 
to the scales $R_1$ and $ R_2$ at the point with coordinates $b_1$ and $b_2$. 
The limits in the double integral are $-\infty$ and $\infty$ for the two 
variables. $\psi $ is the wavelet `mother' function that satisfies the 
constraints
\begin{equation}
\int dx_1\,dx_2\,\psi  = 0,\ \ \ 
\int dx_1\,dx_2\,{\psi }^2 = 1,
\end{equation}

\noindent and the `admissibility' condition (that allows to reconstruct 
the function $f$), i.e. there exists the integral
\begin{equation}
C_{\psi } \equiv (2\pi)^2\int dk_1\,dk_2\frac{{| \hat{\psi} (k_1, k_2)|}^2}
{|k_1\,k_2|},
\end{equation}

\noindent where $\hat{\psi} (k_1, k_2)$ represents the 2D Fourier transform of $\psi$ 
and $||$ denotes the modulus of the complex number.

A reconstruction of the image can be achieved with the inversion formula
\begin{eqnarray}
f(x_1, x_2) & = & \frac{1}{C_{\psi}}\int \frac{dR_1\,dR_2}{{|R_1\,R_2|}^2}
\,db_1\,db_2\,w(R_1, R_2, b_1, b_2) \nonumber \\
& & \times \Psi (R_1, R_2, b_1, b_2; x_1, x_2) \, .
\end{eqnarray}

Next, let us introduce the scalogram of a 2D signal 
\begin{equation}
{\sigma}_w^2(R_1, R_2)\equiv \langle w^2(R_1, R_2, b_1, b_2)
\rangle
\end{equation}
 
\noindent where $\langle \rangle$ means the average value calculated on
the image.

Hereinafter, we shall consider 2D wavelets that are separable, i.e. 
$\psi (x_1, x_2) = \psi (x_1) \psi (x_2)$. In this case, 
${| \hat{\psi} (k_1, k_2)|}^2 = {| \hat{\psi} (k_1)|}^2
{| \hat{\psi} (k_2)|}^2$. In particular, we
are interested in the Haar, the Mexican Hat and the Daubechies 2D-transforms 
that can be generated in terms of the corresponding 1D wavelets. For 
the Haar case, we find
\begin{equation}
{| \hat{\psi} (k)|}^2 = \frac{8}{\pi k^2}{\sin}^4\left(\frac{k}{4}\right), 
\end{equation}

\noindent with an absolute maximum at $k \simeq 4.7$
%satisfying $\tan{\frac{k}{4}}=\frac{k}{2}$
, whereas for the Mexican Hat
\begin{equation}
{| \hat{\psi} (k)|}^2 = \frac{4}{3{\pi }^{1/2}}k^4e^{- k^2}, 
\end{equation}

\noindent with a single peak at $k = \sqrt{2}$. The corresponding formulae 
for the Daubechies wavelets of order $N$ can be found in 
Ogden (\cite{ogden}). The last 
wavelets form an orthonormal basis with compact support,
increasing regularity with $N$ and vanishing moments up to order $N-1$.

Just as an illustration we would like to present the scalogram for a 
CMB signal generated in a Standard Cold Dark Matter (SCDM) model. 
Let us assume that the 
image corresponds to a realization of a random field whose 2-point correlation 
function is homogeneous and isotropic: $\xi (r)$, $r^2\equiv x_1^2 + x_2^2$. 
This is equivalent to assume that the Fourier components 
$\hat{f}(k_1, k_2)$ satisfy
\begin{equation}
\langle \hat{f}(k_1, k_2)\,\hat{f}^*(k_1^{\prime}, k_2^{\prime})\rangle 
= P(k)\delta (k_1 - k_1^{\prime})\delta(k_2 - k_2^{\prime}),
\end{equation}

\noindent where $P(k)$, $k^2\equiv k_1^2 + k_2^2$, is the standard 
Fourier power
spectrum and $\langle \rangle$ means average value over realizations of the 
field (the ergodicity of the field is assumed). 
So, taking average values and using equations (1) and (6), one obtains
the variance ${\sigma}_w^2(R_1, R_2)$ of the wavelet coefficients or scalogram
\begin{equation}
\sigma_{w}^2(R_1,R_2)= R_1R_2\int dk_1\,dk_2\,P(k){| \hat{\psi} (k_1R_1, k_2R_2)|}^2 \, ,
\end{equation}                   

\noindent For 2D white noise, i.e. $P(k) =
constant$, one gets that the scalogram, $\sigma_w^2$, 
is constant at any scale.

On the other hand, if the field $f$ represents the temperature 
anisotropy of the CMB, $\frac{\Delta T}{T}$, one can obtain
\begin{equation}
\left< \left(\frac{\Delta T}{T}\right)^2\right> 
= \frac{1}{C_{\psi}}\int \,dR_1\,dR_2
\frac{{\sigma}_w^2(R_1, R_2)}{R_1^2R_2^2}.
\end{equation}
 
\noindent From the previous equation, ${\sigma}_w^2/C_{\psi}R_1R_2$ represents
the power per logarithmic scale. 
We remark that taking into account the homogeneity and
isotropy of the field, the 2-scale dependence of the scalogram is
redundant in this case. A more appropriate treatment in this continuous 
example would be one
based on isotropic wavelets defined in terms of a single scale. 

\begin{figure}[!t]
\resizebox{\hsize}{!}{\includegraphics{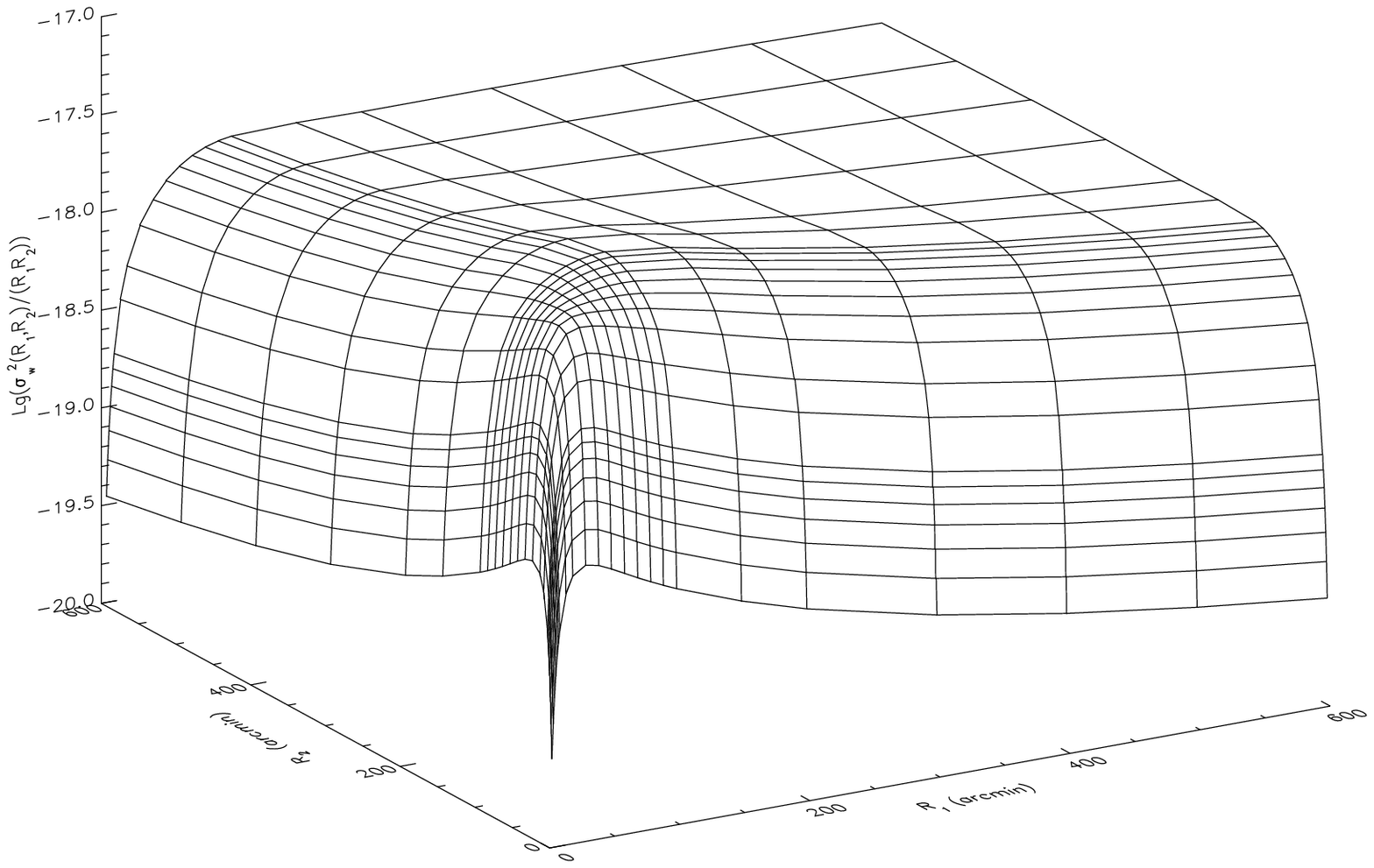}}
\resizebox{\hsize}{!}{\includegraphics{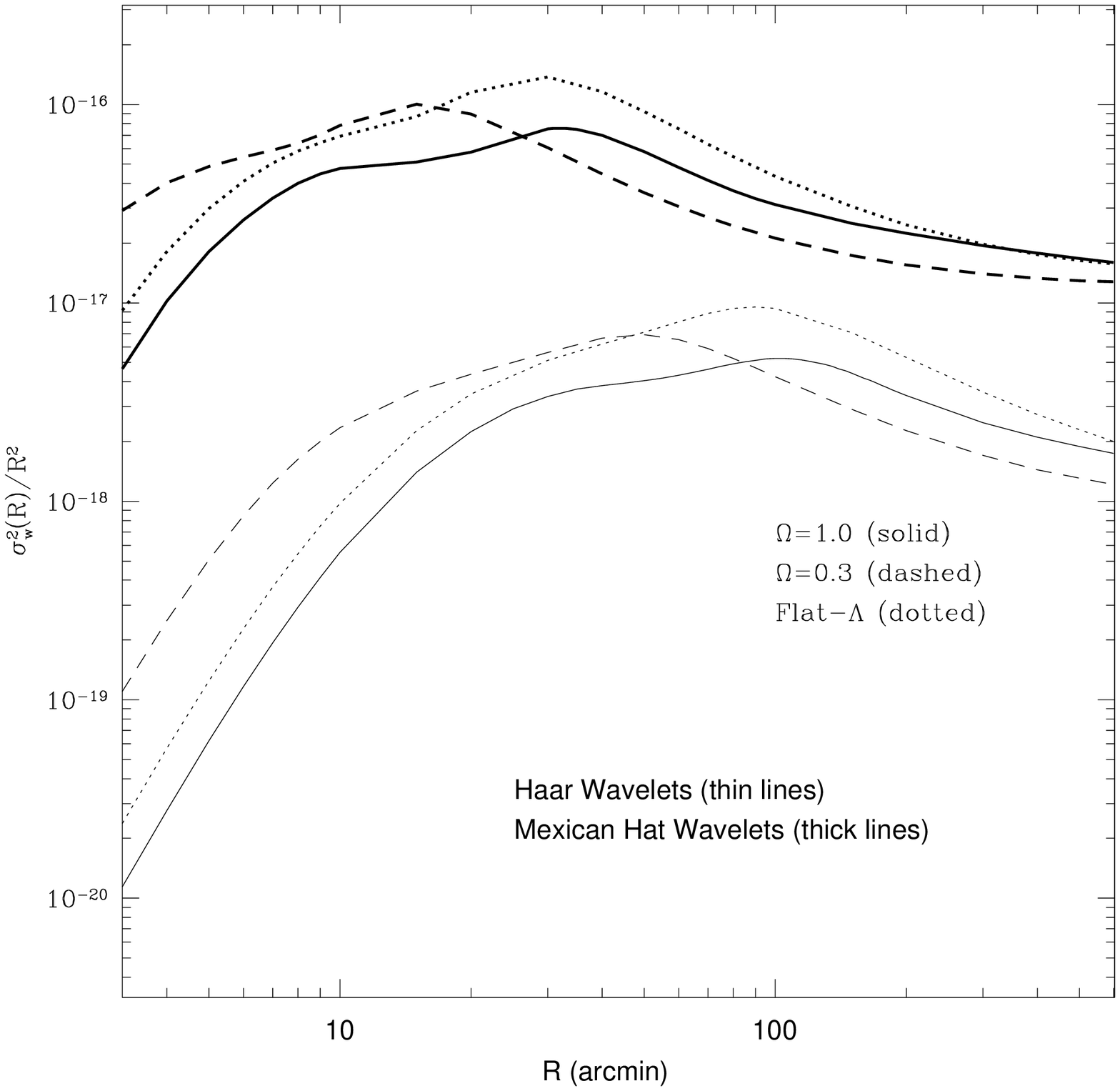}}
\caption{Top: scalogram for a SCDM model using the Haar
wavelet. Bottom: scalogram along the diagonal, i.e., $R_1 = R_2$,
for the SCDM (solid line), open CDM (dashed line) and 
flat-$\Lambda$ CDM (dotted line) models
using the Haar (thin lines, bottom of the panel) and
Mexican Hat (thick lines, top of the panel) wavelets.}
\label{figura1.a}
\end{figure}

In the top panel of Figure~\ref{figura1.a} 
we have represented the scalogram
against the two scales $R_1$ and $R_2$ for SCDM using the Haar transform. The
qualitative behaviour for the other transforms is similar. 
In the bottom panel we compare the scalogram along 
the diagonal for standard, open
($\Omega = 0.3$) and a flat-$\Lambda$ ($\Omega = 0.3 , \lambda = 0.7 $)
CDM models using the Haar and the Mexican Hat transforms. The
qualitative behaviour for the two transforms is similar: there is a plateau for
$R > 1^{\circ}$ and a maximum dependent on $\Omega$,
corresponding to the first Doppler peak. Other secondary
maxima appearing in the Figure are related to the secondary peaks in the 
standard
$C_l$ radiation power spectrum (the $C_{\ell}$ is given by
$C_{\ell}\simeq P(k\simeq \ell )$).
Therefore, the position and amplitude 
of the maxima that appear in the scalogram is model 
dependent, being this quantity tightly related to the $C_l$'s.

%\begin{figure}
%\resizebox{\hsize}{!}{\includegraphics{fig2.eps}}
%\caption{Scalogram along the diagonal, i.e. $R_1 = R_2$,
%for the SCDM (solid line), open CDM (dashed line) and 
%flat-$\Lambda$ CDM (dotted line) models
%using the Haar (thin lines, bottom of the figure) and
%Mexican Hat (thick lines, top of the figure) wavelets.}
%\label{figura1.b}
%\end{figure}

\section{Denoising of CMB maps}

\subsection{2D wavelet method on a grid} 

In general for a grid of $2^n\times 2^n$ pixels, a discretization of the 
parameters of the form: $R_1 = 2^{n - j_1}, b_1 =2^{n - j_1}l_1, 
R_2 = 2^{n - j_2}, b_2 =2^{n - j_2}l_2$ for integer-valued $j$ and
$l$ allows to introduce the 2D discrete wavelet function
\begin{eqnarray}
{\Psi}_{j_1,j_2,l_1,l_2}(x_1, x_2) & = & 2^{(j_1+j_2)/2-n}\psi 
( 2^{j_1 - n}x_1 - l_1) \nonumber \\
& & \times \psi ( 2^{j_2 - n}x_2 - l_2) \, ,
\end{eqnarray}

\noindent where $j_i$ and $l_i$ denote the dilation and the 
translation indexes,
respectively, satisfying $0 \leq j_1 , j_2 \leq n-1, 
0\leq l_1 \leq 2^{j_1}-1, 0\leq l_2 
\leq 2^{j_2}-1$. The resolution level is defined by 
$j = \frac{j_1 + j_2}{2}$, corresponding to 
$R\equiv \sqrt{R_1 R_2} = 2^{n-j}$.
We also introduce a scaling function $\phi$ that allows to define a 
complete basis to reconstruct discrete images,
\begin{equation}
{\Phi}_{0,0,0,0}(x_1, x_2) = 2^{- n}\phi 
( 2^{- n}x_1 )\phi ( 2^{- n}x_2 ),
\end{equation}
\begin{equation}
{\Gamma}^H_{0,j_2,0,l_2}(x_1, x_2) = 2^{j_2/2 - n}\phi 
( 2^{- n}x_1 )\psi ( 2^{j_2 - n}x_2 - l_2),
\end{equation}
\begin{equation}
{\Gamma}^V_{j_1,0,l_1,0}(x_1, x_2) = 2^{j_1/2- n}\psi 
( 2^{j_1- n}x_1 - l_1)\phi ( 2^{- n}x_2 ) ,
\end{equation}

We will consider orthonormal discrete bases as the Haar and Daubechies 
ones. Denoting by $\Lambda$ any of the previous functions, the 
orthonormality condition reads:
\begin{equation}
({\Lambda}_{j_1,j_2,l_1,l_2},{\Lambda}_{j_1^{\prime},j_2^{\prime},
l_1^{\prime},l_2^{\prime}}) = 
{\delta}_{j_1j_1^{\prime}}{\delta}_{j_2j_2^{\prime}}
{\delta}_{l_1l_1^{\prime}}{\delta}_{l_2l_2^{\prime}} ,
\end{equation}

\noindent where $(f,g)$ denotes the scalar product of two functions 
in $L^2(R^2)$. 

The wavelet coefficients are now defined by 
\begin{equation}
w_{j_1, j_2, l_1, l_2} = \int dx_1\,dx_2\,f(x_1, x_2)
\Lambda_{j_1,j_2,l_1,l_2}
\end{equation}

The image is reconstructed using the following expression:
\begin{equation}
f(x_1, x_2) = \sum_{j_1j_2l_1l_2}w_{j_1, j_2, l_1, l_2}
{\Lambda}_{j_1,j_2,l_1,l_2}(x_1, x_2).
\end{equation}

A representation of the wavelet coefficients can
be done by a square that contains small squares and rectangles associated to
different levels of resolution. The first level, representing high-resolution,
is $j_1 = j_2 = 8$ (i.e. $j = 8$) that contains $65536$ wavelet coefficients 
(each one constructed with $2\times 2 = 4$ pixels for the  Haar transform). 
The second level of resolution
contains two boxes: $j_1 = 8, j_2 = 7$ and $j_1 = 7, j_2 = 8$ (i.e. $j = 7.5$)
with a total of $2 \times 32768$ wavelet coefficients 
(each one constructed with
$2^2\times 2 = 2\times 2^2 = 8$ pixels for the Haar transform). 
The levels with $j_1 = 0$ (or $j_2 = 0$) contain both contributions from 
wavelet-wavelet and scaling-wavelet (or wavelet-scaling).
Finally, the lower level of resolution is $j_1 =
j_2 = 0$ (i.e. $j = 0$) and contains four contributions: wavelet-wavelet,
wavelet-scaling, scaling-wavelet and scaling-scaling 
(this last one is proportional to the average value of the image for 
the Haar transform). 
%\begin{figure}
%\resizebox{\hsize}{!}{\includegraphics{fig2.eps}}
%\caption{Spatial distribution of the wavelet coefficients in `boxes' for a
%$32\times 32$ image. The first column and raw correspond to the
%$h_{j_2,l_2}$, $v_{j_1,l_1}$ and $s$ (pixel 1,1) coefficients.} 
%\label{figura2}
%\end{figure}
%A picture of the spatial distribution of the coefficients for an image of 
%$32 \times 32 $ pixels is given in Figure~\ref{figura2}. 

\subsection{Reconstruction of CMB maps and radiation power spectra}

In the present work we have considered simulated maps 
of size $12.8^{\circ} \times
12.8^{\circ}$ square degrees, pixel $1.5^\prime \times 1.5^\prime$ and 
filtered with a $4.5'$ FWHM Gaussian beam for a standard CDM model
($\Omega=1$). 
%We have performed simulations of a standard CDM model ($\Omega=1$).
%Simulations are performed of CDM models with 
%$\Omega = 1$ (standard model), $\Omega = 0.3$ (open case) 
%and a flat-$\Lambda$ model with $\lambda = 0.7$. 
We have included non-correlated Gaussian
noise at different levels (S/N per pixel 
between 0.7 and 3 at the pixel scale), 
considering uniform and non-uniform noise.
This last case is introduced to account for the
non-uniform sampling of satellite observations.
As an extreme case we have simulated a noise map with two different
regions, one with $S/N=3$ (approximately one quarter of the map) and a 
second one with $S/N=0.7$.

The purpose of the denosing of CMB maps is to reconstruct the
original signal map as well as the radiation power spectrum $C_{\ell}$. 

Wavelet decompositions are performed with the 
package {\it 2D-W} developed by our group. 
The procedure uses two scales, $R_1=2^{n-j_1}$,$R_2=2^{n-j_2}$,
$n=9$ in our case. High values of $j=(j_1+j_2)/2$ mean high resolution, i.e.,
small scales. We distribute the coefficients $w_{j_1,j_2,l_1,l_2}$ 
in boxes corresponding to a
couple $(j_1,j_2)$, having a total of $81$ boxes.
The coefficients related to the scaling function
%$h_{j_2,l_2}$, $v_{j_1,l_1}$ and $s$ 
%(corresponding to the first column and raw of 
%Figure~\ref{figura3}) 
are not included in the analysis and they are left untouched.
To perform denoising, the basic operation is the comparison between the 
wavelet
coefficients dispersion of the signal in each box with the one of the noise.
The Gaussian white noise gives the same contribution in all boxes. Since the
signal is negligible at the highest resolutions, the noise dispersion can
be directly estimated from the map. Therefore, the signal dispersion can 
also be estimated for each box. 

\begin{figure}
\resizebox{\hsize}{!}{\includegraphics{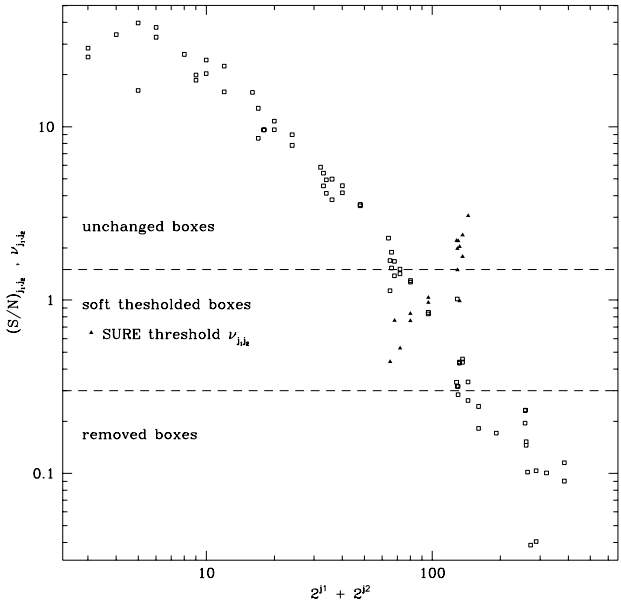}}
\caption{$S/N$ ratio (open squares) in each box for a CMB map 
(SCDM model) with 
uniform noise at the level $S/N = 1$ using Daubechies 4. The three
regions of the plot show the boxes that are kept unchanged 
($S/N \geq 1.5$), 
removed ($S/N < 0.3$) or treated with a soft thresholding technique
(boxes in between). The soft thresholds (solid triangles) 
estimated using SURE are also plotted for the thresholded boxes.}
\label{figura2}
\end{figure}

In Figure~\ref{figura2}, we plot the S/N ratio (defined in
terms of the wavelet coefficients dispersions of signal and noise) for each
box for a CMB simulation with $S/N=1$ in real space.

\begin{figure*}
%\resizebox{\hsize}{!}{\includegraphics{fig7.eps}}
%\resizebox{12cm}{!}{\includegraphics{ds8632sanzf3.eps}}
\hfill
\parbox[b]{55mm}{
\caption{Simulated map of the cosmological signal for the SCDM 
model (top left), signal plus
uniform noise with $S/N = 1$ (top right), denoised map using wavelets 
(middle left) and residual map obtained from the CMB signal map 
minus the denoised
one (middle right). For comparison the denoised map using Wiener 
filter (bottom left) is also shown together with the corresponding 
residuals (bottom right).}
\label{mapas}}
%\label{figura7}
\end{figure*}

\begin{figure*}
%\resizebox{\hsize}{!}{\includegraphics{fig8.eps}}
\resizebox{12cm}{!}{\includegraphics{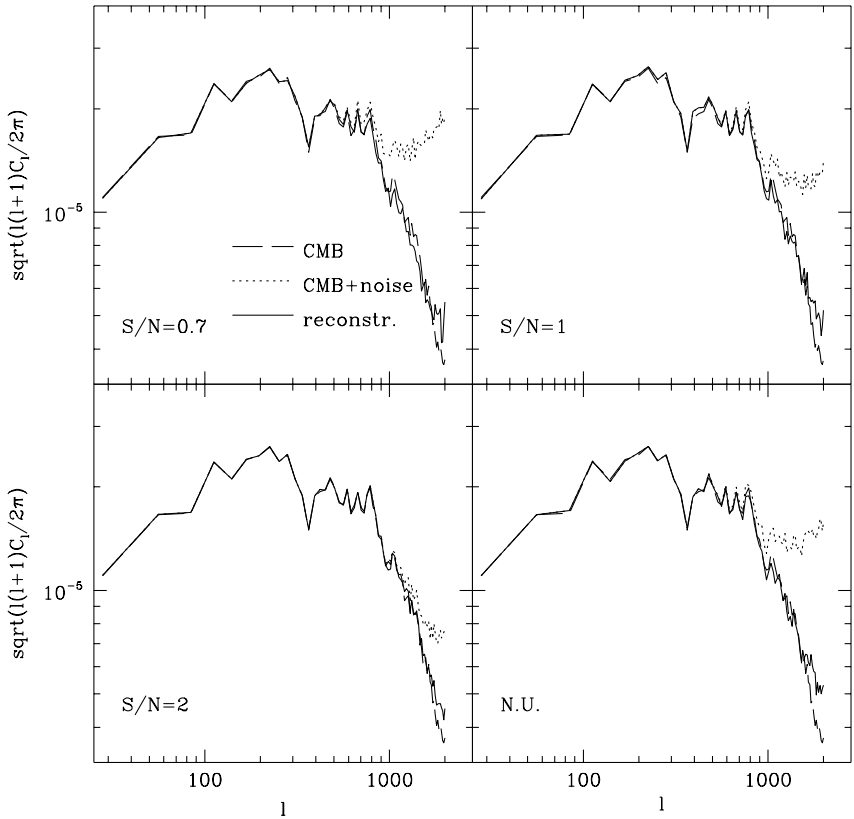}}
\hfill
\parbox[b]{55mm}{
\caption{Power spectrum of the original CMB (dashed line),
CMB plus noise (dotted line) and reconstructed maps using wavelets
(solid line) for different levels of noise.
Top left panel 
corresponds to $S/N = 0.7$, top-right to $S/N = 1$, bottom-left 
to $S/N = 2$ and bottom-right to non-uniform noise 
(the non-uniform noise map consists in two regions of
different S/N ratio; approximately one quarter of the map is at the level
of S/N=3 and the rest at $S/N = 0.7$).}
\label{psreconstruido}}
%\label{figura8}
\end{figure*}

\begin{figure*}
\resizebox{12cm}{!}{\includegraphics{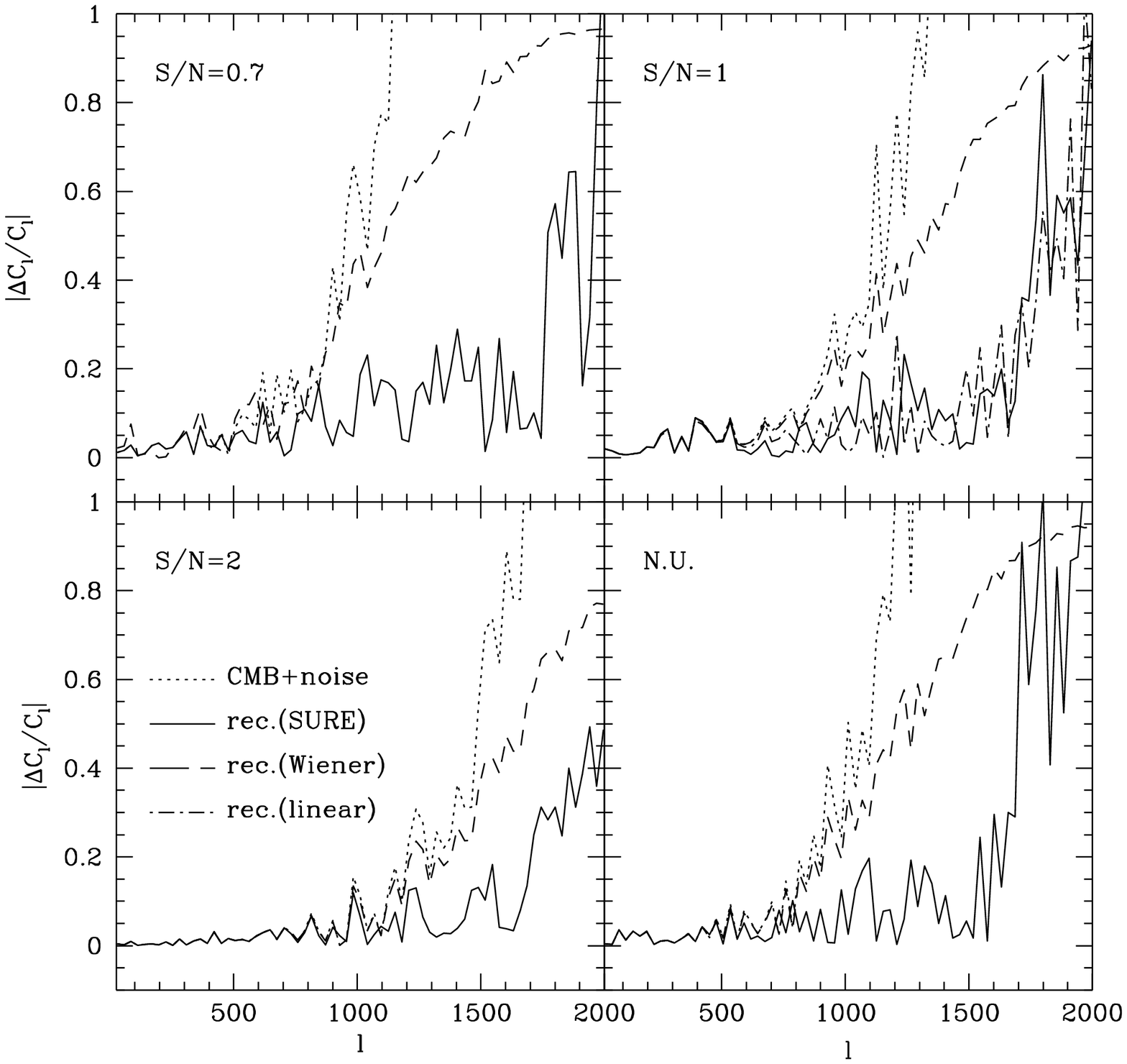}}
\hfill
\parbox[b]{55mm}{
\caption{Absolute value of the relative errors of the CMB power spectrum
obtained from signal plus noise maps (dotted line), wavelet denoised maps
using the SURE thresholding technique
(solid line), Wiener denoised maps (dashed line)
and wavelet denoised map using a linear method, only on the top-right panel
(dot-dashed line). Top-left panel 
corresponds to $S/N = 0.7$, top-right to $S/N = 1$, bottom-left 
to $S/N = 2$ and bottom-right to non-uniform noise.}
\label{relativo}}
\end{figure*}

For the case of uniform noise,
all boxes where the signal dominates ($S/N \geq 1.5$) are kept untouched, 
whereas those with a high level of noise ($S/N < 0.3$) are removed.
On the other hand, the boxes in between are treated with a soft 
thresholding
technique. Given a threshold $\nu$ in terms of the noise coefficients 
dispersion (${\sigma}_n$), the coefficients $|w| > \nu {\sigma_n}$ are 
rescaled as $w^{\prime} = w \pm \nu {\sigma}_n$ (where the $+$,$-$ 
signs correspond 
to negative and positive values of $w$ respectively), 
whereas the remaining 
coefficients are set to zero. The threshold $\nu$ for each box is chosen 
using the SURE 
method (Donoho \& Johnstone \cite{donoho}). 
%This is a signal
%independent thresholding technique, defined under the assumption 
%of uniform white Gaussian noise. 
The threshold is obtained by
minimization of an unbiased estimate of the expected mean squared 
error of the estimation of the
signal wavelet coefficients (see for instance Ogden \cite{ogden}). 
In Figure~\ref{figura2}, 
the thresholds obtained with the SURE technique are plotted for a
CMB map with $S/N = 1$. As expected, lower 
S/N levels are treated with higher thresholds, i.e., more coefficients
are removed.
Changing the range of S/N where the soft technique 
is applied (providing is around $S/N = 1$), does not appreciably change 
these results.
We have used Daubechies 4 but we obtain little or no 
variations if we adopt higher order Daubechies wavelets.
However, the Haar transform gives worse results. 
Table 1 shows the error in the map reconstructions for different 
$S/N$ ratios with Gaussian uniform noise.
The error improvement achieved with the denoising
technique applied goes from factors of $3$ to $5$ for $S/N = 3 - 0.7$.
The four top panels of Figure~\ref{mapas} show CMB maps with only signal 
(SCDM), signal plus noise with a $S/N = 1$, the reconstructed map 
using wavelets and the residual one.

\begin{table}
%\begin{center}
\caption{Reconstruction errors (\%)}
\label{errores}
\begin{tabular}{|c|c|c|}
\hline
S/N & SURE & linear \\
\hline
0.7 & 27.4 & 29.4 \\
1.0 & 21.7 & 23.4 \\
2.0 & 13.3 & 14.4 \\
3.0 & 10.0 & 11.1 \\
N.U.& 24.3 &  -- \\
%N.U.& 24.3 & 26.1 \\
% en lineal: 25.4 si el threshold se hace por pixel
%N.U.$^1$& 25.1 \\
\hline
\end{tabular}
%\end{center}          
\end{table}

%\begin{table}
%\caption{Reconstruction errors}
%\label{errores}
%\begin{tabular}{@{}cc@{}}
%\hline
%\hline
%S/N & Image error($\%$) \\
%\hline
%0.7 & 27.4 \\
%1.0 & 21.7 \\
%2.0 & 13.3 \\
%3.0 & 10.0 \\
%N.U.& 24.3 \\
%N.U.$^1$& 25.1 \\
%\hline
%\end{tabular}
%\begin{list}{}{}
%\item[$^1$] non uniform noise is assumed (see text)
%\end{list}             
%\end{table}

Regarding non-uniform (N.U.) noise, wavelet techniques 
allow us to treat each location in the image separately.
At each fixed location and fixed $(j_1,j_2)$ we calculate the 
dispersion of the corresponding noise wavelet coefficient
from 500 simulations of our non-uniform noise.
Since we consider non-uniform noise that is uncorrelated,
%the considered non-uniform noise is uncorrelated,
the average noise dispersion is the same for all the boxes,
as in the uniform noise case.
Therefore, we can get again the dispersion of the signal for each 
$(j_1,j_2)$ pair as well as 
the S/N ratio for each coefficient. Those coefficients with 
S/N ratio in the
considered range ($ 0.3 \le S/N < 1.5 $) are treated with a soft 
thresholding technique, whereas the rest are either kept or removed 
depending on their S/N ratio.
Since in presence of non-uniform noise, we cannot use the SURE
technique to estimate the optimal threshold $\nu$ (as far as we know,
work is in progress to define a general threshold in the case of 
non-uniform noise, Von Sachs \& McGibbon \cite{vonsachs}),
we choose for all the thresholded coefficients $\nu = 1$. 
This threshold is defined with respect to the noise
dispersion in each particular coefficient.
We have chosen this value of $\nu$ because it gives 
a good reconstruction when comparing with the original map,
but the results are not very sensitive to the choice of a
different threshold in the range $\nu = 0.8 - 2.0 $.
In Table 1 we present the error of the reconstructed map in the 
presence of non-uniform noise as considered in this work. 

%Of course, to apply this scheme, a good knowledge of the statistical 
%properties of the noise is needed. On the other hand, this may not be the 
%situation in some experiments. In this case, we can proceed assuming 
%that we have uniform noise (with dispersion equals to the average
%dispersion at the highest resolution level). 
%This procedure can be seen as a different choice of the 
%parameters (S/N range to apply the soft thresholding and the corresponding 
%soft thresholds) when applying the technique coefficient by coefficient.
%Since the method is weakly
%dependent on the chosen parameters, for a realistic case of
%non-uniform noise, the reconstruction will get only slightly worse.
%In Table 1 we present the error of the reconstructed map in the 
%presence of non-uniform noise as considered in this work. 
%As can be seen the
%error in the image increases by a $ \sim 1\% $ when uniform noise 
%is assumed.

%\begin{figure*}
%\resizebox{\hsize}{!}{\includegraphics{fig9.eps}}
%\caption{Absolute value of the relative errors of the CMB power spectrum
%obtained from signal plus noise maps (dotted line), wavelet denoised maps
%(solid line) and Wiener denoised maps (dashed line). Top-left panel 
%corresponds to $S/N = 0.7$, top-right to $S/N = 1$, bottom-left 
%to $S/N = 2$ and bottom-right to non-uniform noise.}
%\label{figura9}
%\end{figure*}

Regarding the power spectrum, $C_{\ell}$, the denoising method performs very
well. Figures~\ref{psreconstruido} and ~\ref{relativo} show the 
reconstructed spectrum and the relative error
for $\ell < 2000$ for different $S/N$ ratios
(the power spectrum is obtained in the usual way, 
averaging over the Fourier modes of the considered map at each $k$). 
The relative 
error is $\mathrel{\hbox{\rlap{\hbox{\lower4pt\hbox{$\sim$}}}\hbox{$<$}}}$
$20\%$ for $\ell < 1700$ in all the considered cases except for the map 
with $S/N = 0.7$. In this last case the error increases to 
$\mathrel{\hbox{\rlap{\hbox{\lower4pt\hbox{$\sim$}}}\hbox{$<$}}} 30 \%$ 
for the same range of $\ell's$.
%Since the power spectrum of the noise is constant at each scale, 
%one could also recover the $C_\ell$'s of the original signal by 
%subtracting
%the power spectrum of the signal plus noise map minus the estimated
%power spectrum of the noise. However, this method gives in general 
%worse results than the wavelet techniques, although better than 
%Wiener Filter.

%Notice that Wiener Filter is the optimal linear filter to denoise a map
%regarding the reconstruction error of the image. However, it is well 
%known that this method underestimates the power spectrum at
%small scales giving higher relative $C_\ell$ errors as can be
%seen from figure Figure~\ref{relativo}.

%\begin{figure}
%\resizebox{\hsize}{!}{\includegraphics{fig10.eps}}
%\caption{$W_j$ for the cosmological signal (dashed line), 
%uniform Gaussian noise with $S/N = 1$ (dashed-dotted line),
%signal plus noise (dotted line) and reconstruction (solid line)
%using Daubechies 4.}
%\label{figura10}
%\end{figure}

Finally, we have looked for possible non-Gaussian features introduced by the 
non-linearity of the soft thresholding technique. 
We have compared the probability density
function of the original and the reconstructed maps 
using a Kolmogorov-Smirnov test. 
Both distributions are compatible at similar or higher levels than
the original map and the corresponding linear reconstruction 
obtained with the Wiener Filter technique.
Moreover, not significant change is observed in the skewness and
kurtosis of the original and reconstructed maps. 
However, the application of soft thresholding to the 
wavelet coefficients at a certain level, which are Gaussian distributed
for a Gaussian temperature field,
clearly changes their
distribution by removing all coefficients whose absolute values
are below the imposed threshold and shifting the remaining ones
by that threshold. Therefore, this technique is introducing a certain
level of non-Gaussianity that will depend on the threshold imposed and
that should be taken into account when analysing the data.
If we are mainly concern about preserving the Gaussian character
of the reconstructed image, denoising with wavelet techniques is
still possible.
%,rather recovering the power spectrum of
%the image itself, we can still do it with wavelets.
Instead of using a soft thresholding technique, we can apply a
linear denoising method in wavelet space. In particular, we have
removed all wavelet coefficients at boxes with $S/N < 1$ and 
left the rest untouched. 
%In the case of non-uniform noise, we have
%proceed assuming that we have uniform noise with dispersion equals
%to the average dispersion at the highest resolution level.
The reconstruction errors get only slightly worse with this 
simple linear
technique as shown in Table~\ref{errores}. Regarding the reconstructed
power spectrum, this is at the same level than the SURE reconstruction
(see top-right panel of figure~\ref{relativo}) for all the 
considered cases. It is important to remark that the linear 
denoising method based on 2D wavelets performs much better than the 
Multiresolution one due to the larger number of boxes corresponding
to the product of the 2 scales considered. 

\subsection{Comparison with other denoising methods}

A comparison between Wiener Filter (see for instance 
Press et al. \cite{press}) 
and wavelet techniques has also been performed.
In relation to map reconstruction the error affecting the Wiener
reconstructed maps is comparable to that achieved with wavelet
techniques in all the considered cases. However, in order to apply 
Wiener filter previous knowledge of the signal power spectrum is required. 
In a real situation this may well not be possible.
The reconstructed and residual maps using Wiener Filter are shown in 
Figure~\ref{mapas}. 
In addition, when using Wiener Filter, the power
spectrum of the reconstructed image is clearly suppressed at high $\ell's$,
giving much worse results than the
wavelet technique. For comparison, we have plotted in 
Figure~\ref{relativo} the 
absolute value of the relative error of the $C_{\ell}'s$ for the 
reconstructions with Wiener Filter.
On the other hand, one could recover the $C_\ell$'s of the original 
signal by subtracting from
the power spectrum of the signal plus noise map the estimated
power spectrum of the noise, which is constant at each $\ell$.
However, this method gives in general 
worse results than the wavelet techniques
and besides cannot be used to reconstruct the image. 
We have also applied a Maximum Entropy Method to the maps used in this
work, with the definition of entropy given by Hobson \& Lasenby 
(\cite{hobson98b}).
This method provides reconstruction errors at similar levels as 
the wavelet techniques. However, we remark that wavelet
techniques are computationally faster 0(N) and simpler to apply 
(not requiring iterative processes) than Maximum Entropy Methods.

\section {Conclusions}

We have considered the 2D wavelet transform with two scales to analyse CMB 
maps. First of all we present the continuous approach 
to the 2D wavelet.
A discrete analysis is performed for a finite image. In this
case, a scaling function is introduced in order to define a 2D wavelet basis.

The 2D wavelet technique has been applied to denoise CMB maps for 
S/N ratios ranging between $0.7$ and $3$. We have also considered 
the case of non-uniform Gaussian noise. A factor between $5$ and $3$ is 
gained for the reconstructed images / original
signal map in relation
to the signal plus noise maps / original signal map.
Regarding the $C_{\ell}'s$, the relative
errors are below a $20\%$ up to $\ell = 1700$ for all the cases with
S/N $\geq 1$. A comparison with Wiener Filter and Maximum Entropy Method 
has also been performed.
The later gives comparable reconstructions to the wavelet method,
being however slower and more complicated to apply.
Wiener Filter provides reconstructed maps with errors comparable to 
the wavelet 
technique we propose. However, unlike the wavelet method, Wiener filter
requires previous knowledge of the signal power spectrum. Moreover,
the $C_\ell's$ of the denoised map obtaining by applying Wiener filter 
are clearly underestimated.

Finally, we would like to remark that linear wavelet denoising, which 
preserves Gaussianity, gives reconstruction errors similar to
those obtained with the non-linear soft thresholding techniques.
Moreover, linear denoising is better achieved by 2D wavelets than
by the multiresolution ones discussed in
Sanz et al. (\cite{sanz}).

\begin{acknowledgements}  
This work has been supported by the DGESIC Project
no. PB95-1132-C02-02, CICYT Acci{\'o}n Especial no. ESP98-1545-E and 
Comisi{\'o}n Conjunta Hispano-Norteamericana de Cooperaci{\'o}n
Cient{\'\i} fica y Tecnol{\'o}gica with ref. 98138. RBB has been 
supported by a Spanish MEC fellowship. JLS acknowledges partial financial 
support from a Sabbatical Grant  by the Spanish DGESIC. JLS, RBB and
LC thank the hospitality of the Center for Particle Astrophysics and
the Astronomy Department of the University of California at Berkeley.
LT acknowledges partial financial support from the `Agenzia Spaziale
Italiana' (ASI) and Consiglio Nazionale delle Ricerche (CNR).
\end{acknowledgements}

\end{document}